\documentclass[11pt,journal]{IEEEtran}

\usepackage{color,xcolor}
\usepackage[pdftex]{graphicx} 
	\graphicspath{{./}}

\usepackage[colorlinks=true,linkcolor=blue,citecolor=blue,hyperfootnotes]{hyperref}

\usepackage{array}
\usepackage{multirow}

\begin{document}
\IEEEoverridecommandlockouts

\thispagestyle{empty}

\begin{figure*}[!t]\large
This paper is a preprint; it has been accepted for publication in IEEE Consumer Electronics Magazine (DOI: 10.1109/MCE.2019.2892221).
\medskip

{\bf IEEE copyright notice}
\smallskip

\copyright\ 2019 IEEE. Personal use of this material is permitted. Permission from IEEE must be obtained for all other uses, in any current or future media, including reprinting/republishing this material for advertising or promotional purposes, creating new collective works, for resale or redistribution to servers or lists, or reuse of any copyrighted component of this work in other works.
\vspace*{300pt}

\mbox{~}
\end{figure*}

\newpage

\title{Secured by Blockchain: Safeguarding Internet of Things Devices}

\author{Nicholas Kolokotronis, \and Konstantinos Limniotis, \and Stavros Shiaeles \and and \and Romain Griffiths}

\maketitle

\begin{abstract}
Blockchain is a disruptive technology that has been characterised to be the next big thing and has already gained a broad recognition by experts in diverse fields. In this paper, we consider possible use cases and applications of the blockchain for the consumer electronics (CE) industry and its interplay with the Internet of things. Instead of discussing how the blockchain can revolutionise the supply chain, we focus on how it could be employed for enhancing the security of networked CE devices. This work is motivated by the large number of recent attacks that use easily hackable devices as a weaponry. Towards this direction, privacy and data protection aspects of blockchain solutions are also presented and are linked to regulatory framework provisions. Information on existing blockchain solutions is also provided.
\end{abstract}

\section{Introduction}
\label{sec:intro}

\IEEEPARstart{T}{he} vision of the {\em Internet of things} (IoT) is to establish a whole new ecosystem that is comprised of heterogeneous connected devices communicating to deliver environments that make the way we do business, communicate, and live far more intelligent \cite{corcoran16}. In the following years, almost anything in the surrounding environment will be interconnected with billions of other devices, as part of a network of networks. Examples of IoT devices include sensors and embedded devices in buildings, industrial control systems, etc., as well as, {\em consumer electronics} (CE) devices, like digital cameras, TVs, computers, and smartphones \cite{thapliyal18}.

\begin{figure}[t]
\centering
\includegraphics[scale=0.525]{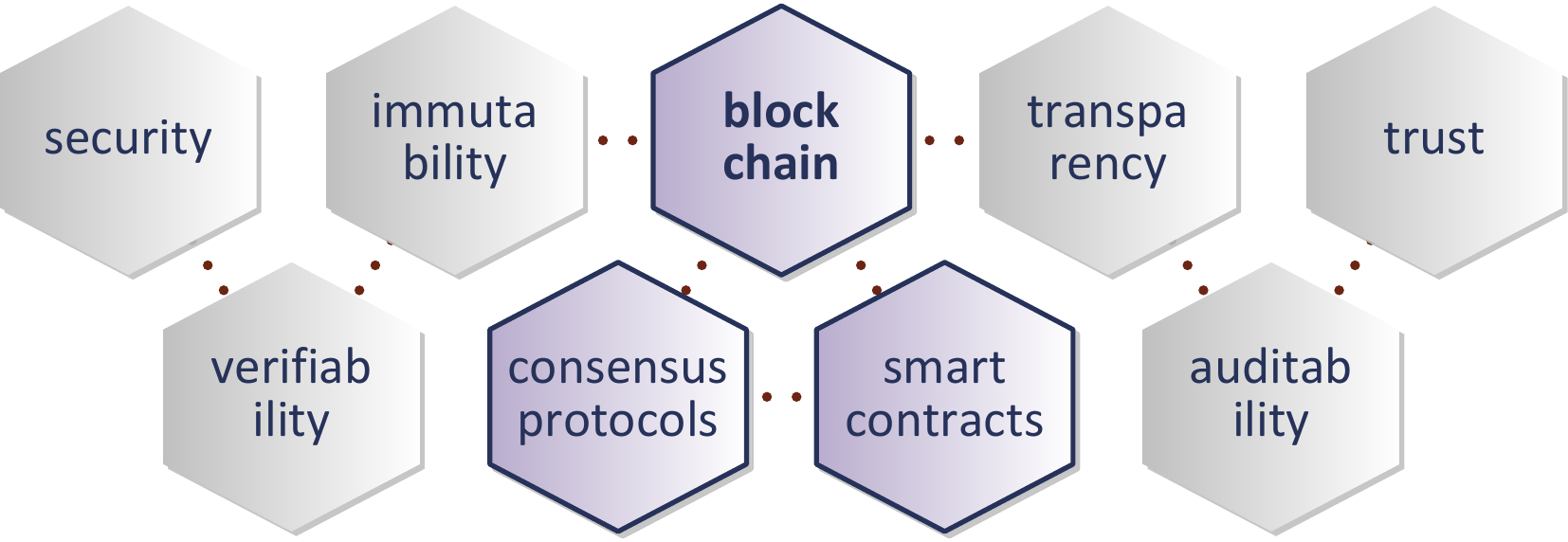}
\caption{Main advantages of blockchain/distributed ledger technology.}
\label{fig.adv}
\end{figure}

The technological and industrial revolution brought by the IoT could be amplified if combined with blockchain solutions \cite{christidis16}. The blockchain, which is the data structure underlying the Bitcoin, provides a verifiable process for storing transactions or digital assets, on an immutable shared ledger, in a way that it is transparent, secure, and robust ({\em see} Fig. \ref{fig.adv}); every transaction is accompanied by an auditable proof that is valid and has been accepted and mutually agreed by the nodes. The adoption of blockchain, or {\em distributed ledger technology} (DLT), in IoT would allow devices to act autonomously and execute transactions via smart contracts. Thus, beyond its use in cryptocurrencies, the blockchain has the potential to impact other industries, like healthcare and CE \cite{lee17a}.

The above technological evolution comes with new forms of threats or attacks that exploit the complexity and heterogeneity of IoT networks, therefore rendering security amongst the most important aspects of a networked world \cite{lee17b}. The fact that the number of intelligent things has greatly increased in the last few years, and will continue to do so, amplifies concerns about the security of networked devices, applications, and services. These often constitute the target of attackers, since they may easily exploit well-known vulnerabilities to accomplish their objectives, e.g. gain unauthorised access to the device, steal sensitive data, deny services to legitimate users, and use it as a vehicle to launch other advanced attacks. Thus, there is an urgent need for securing the communications among untrusted devices to allow them establish trust and operate transparently.

In this paper, we investigate whether the blockchain could enhance the security of IoT-enabled CE devices in a cryptographically verifiable manner. In particular:
\begin{itemize}
\item we further explore novel security and privacy applications of the blockchain in the IoT domain, which have recently been recognised \cite{puthal18a};
\item we illustrate how cyber-security attacks tampering CE devices' critical files can be mitigated using the blockchain as an independent root-of-trust;
\item we describe how integrity data on the blockchain can be leveraged to define a trust-based framework for coordination between IoT devices;
\item we link blockchain solutions' privacy and data protection aspects to regulatory framework provisions.
\end{itemize}
The above new paradigm, driven by the blockchain, can bring the transparency and auditing needed for trusting online services without a {\em trusted third party} (TTP). 

Although the potential applications of blockchain in the IoT are extensively explored, the area of using it for strengthening IoT security and privacy or addressing cyber-security needs is still in its infancy. Many challenges remain to be tackled so as to build blockchain-based solutions for the IoT, including processing power, storage, and scalability. The fact that data stored on the blockchain cannot change and are public, raises even more challenges: data confidentiality, the need for long-term security, and the right to be forgotten.

\section{How the blockchain works}
\label{sec:blockchain}

The blockchain was introduced with Bitcoin as part of a solution to tackle, in a distributed fashion, the double-spending problem in a trustless network of peers. The solution relies on cryptographic mechanisms that ensure the immutability of data stored on the ledger; a {\em security through transparency} approach is taken, according to which all nodes' transactions are public, and thus anyone can verify their validity. The transactions are digitally signed with the owner's {\em private key} and are verified with the associated {\em public key}. New transactions are packed into blocks, containing links to past transactions (thus creating a chain of blocks), and they are subsequently appended to the structure. The ledger's maintenance is carried out by the nodes called {\em miners} \cite{puthal18b}. The mutual agreement on the validity of newly created blocks is done according to a {\em consensus} protocol.

\subsection*{Blockchain and IoT considerations}

The design of a blockchain solution for securing IoT devices is not trivial. In most cases, a device's resources are highly constrained, whilst there is a need for performing transactions at high speed. These requirements call for efficient blockchain solutions; key design factors that determine both their security and performance, in the context of the IoT, are briefly presented below.

\subsubsection*{Modelling}

Depending on whether the ledger is open to the public, i.e. it can be used by all network nodes, it is classified as public or private. Moreover, if the miners that maintain the ledger have been selected a priori, then the ledger is called permissioned; otherwise, if any node can be a miner, the ledger is said to be unpermissioned.

In an IoT security scenario, the blockchain that should be designed needs not necessarily be universal; in fact, there may be many local and global blockchains with different purposes; the use of sidechains could also prove to be efficient in certain cases. The model to be used in each case depends on security, scalability, performance, and other critical for the IoT scenario requirements. There are trade-offs between the above criteria: a private blockchain with less users could minimise the integrity verification time and enjoy almost immediate tamper resistance and detection; on the other hand, this choice reduce security, since we rely on less nodes to maintain the data structure.

\subsubsection*{Consensus}

There are many consensus protocols that have been proposed. Their goal is to allow nodes agree on a {\em single} version~of valid transactions. The {\em proof of work} (PoW) and {\em proof of stake} (PoS) protocols are the most prominent examples; {\em see} e.g. \cite{puthal18b} for more details. The processing power that PoW consensus algorithms require to be devoted can be adjusted to the application needs in order to meet performance requirements. It is clear that lowering the hardness of the computational puzzle to solve also impacts the security offered. Hence, an optimal balance should be found if PoW is to be used in blockchain-based IoT applications.

\subsubsection*{Smart contracts}

These are computer scripts that are stored in, and are automatically executed by, a distributed ledger once they are triggered \cite{christidis16}. They are an important part of blockchain-based IoT applications, where IoT devices are expected to be highly autonomous and transact based on some predefined criteria.

\section{Need for securing IoT-enabled devices}
\label{sec:issues}

Security and privacy are increasingly important factors for the acceptance of IoT products and services. There are many recent attacks that exploit IoT devices to perform {\em distributed denial of service} (DDoS) attacks, to spy on people, and hijack communication links, delivering full control of anything that is remotely accessible to an attacker. Recent reports indicate that DDoS, cloud-based, and mobile attacks are among the most common attacks. The availability of botnets-for-hire has led to the noticeable increase in DDoS attacks and it is highly likely that the IoT will further facilitate the formation of such botnet armies. A recent example of DDoS attack in Oct.\ 2016, attributed to Mirai botnet, affected millions of users and companies, also crippling the servers of popular services, like Twitter, Netflix, and PayPal; this simple malware infected the IoT devices that used default settings and credentials.

Most of the security issues arise from devices with flawed design or poor configuration, which allows attackers to  easily compromise them \cite{decuir15}. Tools such as {\em Shodan} and {\em IoTseeker} can be easily employed to discover such vulnerable devices. This brings the important question of how can large-scale exploitation of such vulnerabilities be prevented, as IoT devices have very limited capacity for securing themselves; they cannot be equipped with the operating systems or the multitude of security mechanisms available on a desktop computer. Moreover, a software update method to fix vulnerabilities and update configuration settings is often overlooked by manufacturers, vendors, and others on the supply chain. Further, even if such functionality is given, there is often no efficient way to patch those devices, and the possibility to add new vulnerabilities exists.

Many best practices have been developed in order to address these issues. As an example, the {\em online trust alliance} (OTA) published an IoT trust framework for the CE devices, whose recommendations have technical counterparts that have been widely recognised to be the cornerstone towards securing the IoT. Among these security solutions, the following are priority controls to implement for enhancing attack prevention:

\begin{itemize}
\item manage efficiently the hardware devices;
\item develop an inventory of authorised software;
\item protect the configurations of CE devices;
\item perform continuous vulnerability assessment;
\item protect sensitive data and users' privacy.
\end{itemize}
Building and managing vulnerability profiles, possibly with the involvement of manufacturers \cite{kshetri17}, could assure consumers that security and privacy issues are addressed seriously. Realising the above is far from trivial and blockchain may prove to be ideal in this direction.

\section{Placing trust on the blockchain}
\label{sec:trust}

Current centralised security solutions are not adequate for dealing with the waves of attacks and the heterogeneity of the IoT devices.  The following analysis leads to the conclusion that blockchain can be used to achieve trusted decentralised coordination among IoT devices and help defending against sophisticated attacks. It is expected to define a fundamentally new approach to security, going beyond the device's security alone, to include \cite{kshetri17}:

\begin{itemize}
\item {\em Identity security:} blocking identity theft, disallowing successful use of rogue public-key certificates, countering man-in-the-middle (MiTM) attacks.

\item {\em Data security:} preventing data tampering, developing access control mechanisms and {\em keyless signature infrastructures} (KSI) on the blockchain.

\item {\em Communication security:}, protecting {\em domain name system} (DNS) services, stopping DDoS attacks, defending critical information infrastructures.
\end{itemize}
Specifically, the {\em security through transparency} approach of a public blockchain has clear advantages for the IoT compared to the usual {\em security through obscurity} model.

\begin{figure}[t]
\centering
\includegraphics[scale=0.525]{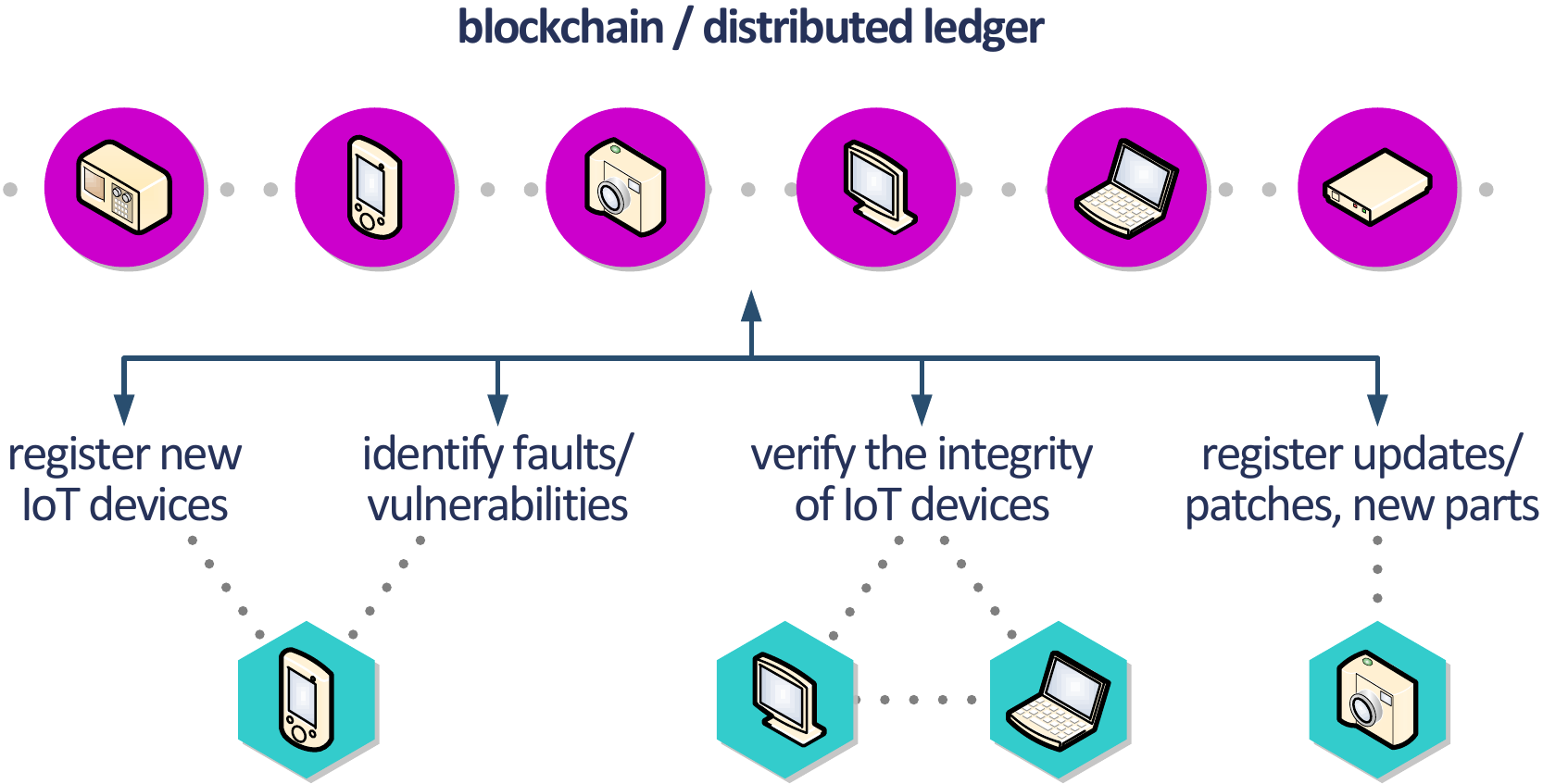}
\caption{The blockchain functions as a distributed ledger, for various IoT transactions enhancing the security of IoT-enabled CE devices.}
\label{fig.sec}
\end{figure}

\subsection*{Can devices' integrity be protected?}

Attacks on connected smart devices aim at impacting their operational integrity so that they do not strictly function within their specified usage. For lightweight devices, lacking proper defence mechanisms, critical information of a manufacturer's IoT device operation could be recorded on the blockchain so that it can be later queried ({\em see} Fig. \ref{fig.sec}) when e.g. a verification of proper functioning is needed, or when software parts have to be updated reliably \cite{christidis16}. Thus, multiple aspects, such as
\begin{itemize}
\item device's firmware;
\item operating system and critical software;
\item system/network configuration files;
\item audit and event logs;
\end{itemize}
could be verified against a history of previously valid states, to ensure their integrity. Such information could be monitored on a continual basis for illicit changes. This approach fits well within the current practices of the software distributors that publish their software binaries' hash to allow users verify the authenticity of their copy \cite{melara15}. Although preserving integrity alone cannot possibly thwart attacks targeting at other security properties, like availability, it sets the ground for establishing a trusted execution environment for the implemented security controls and services. To enhance security in IoT-enabled CE devices, the following phases during their life-cycle should be considered~as shown in Fig. \ref{fig.sec}.

\subsubsection*{Registration}

When assembled, a product is registered into a blockchain, linking its cryptographic fingerprint to an entry in the blockchain.

\subsubsection*{Update}

Upon change, e.g. firmware update, a new fingerprint is generated and submitted to the network of peers who will insert the fingerprint into their local copies via a consensus algorithm.

\subsubsection*{Verification}

At any point, nodes can verify a device's properties by regenerating the fingerprint and comparing this value against the (correct) entry in the blockchain.

\subsection*{Can devices establish mutual trust?}

As already noted in \cite{christidis16}, information on the blockchain can be leveraged to allow devices establish trust relations that are achieved as an emergent property from their interactions. In principle, trust is a complex notion related to the belief on the security, integrity, reliability, and other aspects of a device. There is no universal definition of trust, but it can be perceived as the expectation that a device will behave correctly for some specific purpose and will pose no threats or risks to any other parties involved; both {\em objective} (e.g. vulnerabilities' exposure, integrity status, etc.) and {\em subjective} (e.g. recommendations or reputations) measurements contribute to its computation \cite{yan14}. Focusing on security and integrity aspects of IoT devices, one could consider the following objective measurements towards evaluating a device's trust score:

\begin{itemize}
\item Have critical files or firmware been tampered with?
\item Have the latest software patches been installed?
\item Is the IoT device exposed to known vulnerabilities?
\item Is the network traffic being generated typical?
\end{itemize}
To accurately answer the above, the state-of-the-art in many security areas has to be combined. Implementing controls for monitoring a device's behaviour so as to protect users' privacy is a significant open problem for IoT-enabled CE devices. The {\em manufacturer's usage description} (MUD) specifications can be adopted for enforcing operational usage compliance and block suspicious connections or services.

Practices that utilise such information in a general reputation-based setting are ideally combined with blockchain solutions due to their transparency and their ability to be regulated by the whole network of peers. This approach is along the same lines with the implementation of CE device {\em blacklisting}, but much more sophisticated than that, which has been suggested as the protection means by network operators against devices being ultimately untrusted (e.g. stolen mobile phones). Likewise, blockchain solutions, by relying on smart contracts, can facilitate the wide adoption of such practices.

\subsection*{How secure is the blockchain?}

Research on blockchains' potential applications in the security area has been growing. There have been proposals for using blockchain in the form of cryptocurrencies alternative to bitcoin (they are called {\em altcoins}) or as the core structure accompanied by some application-tailored consensus protocol. Examples include decentralised access-control management systems, where users own and control their personal data, binary and certificate transparency systems \cite{melara15}, and cryptocurrencies to allow a device proving having contributed to a DDoS attack against a specific target. The security of these proposals, wherever rigorously treated, depends on the assumptions made about the security of the underlying blockchain data structure. However, it is now well-understood that a holistic security analysis must consider cryptographic (primitives employed), software (smart contracts), and game-theoretic (incentives) aspects.

\begin{table}[t]
\setlength{\tabcolsep}{3pt}
\caption{Malicious attacks to blockchain and defenses \cite{xu16}.}
\label{tab.att}
\centering
\begin{tabular}{@{}p{1.25cm}p{4.25cm}p{2.95cm}@{}}
\hline
Attack             & Definition & Defensive measures \\
\hline
Double spending    & Many payments are made with a body of funds & Complexity of mining process  \\
Record hacking     & Blocks are modified -- fraudulent transactions are inserted & Distributed consensus\\
51\% attack	       & Miner with more than half of the network's computational power, dominating verification process & Detection methods and design of incentives \\
Identity theft     & An entity's private key is stolen & Reputation blockchain on identities \\
System hacking     & Software systems that implement a blockchain are compromised &	Advanced intrusion detection systems \\
\hline
\end{tabular}
\end{table}

From the cryptographic viewpoint, blockchain's properties have been well-studied due to the attention gained by Bitcoin. {\em Persistence} and {\em liveness} are critical properties for blockchain security, i.e. to prevent adversaries from performing a selective DDoS attack against account holders or mining pools; it is known that these cannot hold if more than $1/2$ of the miners in a synchronous network are selfish (i.e. they do not follow the protocol) ---known as the 51\% attack. This threshold has been subsequently revised, using a game-theoretic approach, to letting an adversary's hashing power be less than about $1/3$ of the network's total hashing power. Since the assumption on fully synchronous networks (absence of any delays in message delivery) that is often made is unrealistic, research focuses on asynchronous networks to study  blockchain solutions' security. A synopsis of main  attacks is given in Table \ref{tab.att}.

Although it is hard to modify data in a blockchain, it is possible to compromise software systems implementing the technology; the hack of Mt.\,Gox, resulting in \$450 million losses, is such a notable example. Another incident is related to the {\em decentralized autonomous organization} (DAO), holding a large percentage of Ether; it suffered $\sim$\$60 million in losses when a smart contract vulnerability was exploited that blocked the invocation of the function updating a user's balance; a summary of such vulnerabilities in smart contracts is provided in Table \ref{tab.smart}. Many of these vulnerabilities apply to Solidity, the high-level language supported by Ethereum. 

\begin{table}[t]
\setlength{\tabcolsep}{3pt}
\caption{Taxonomy of vulnerabilities in smart contracts \cite{li17}.}
\label{tab.smart}
\centering
\begin{tabular}{@{}p{2.4cm}p{4.55cm}p{1.5cm}@{}}
\hline
Vulnerability        & Cause                                 & Level \\
\hline
Call to unknown      & The called function doesn't exist     & Contract's \\
Out-of-gas send      & Fallback of the callee is executed    & source code \\
Exception disorder   & Exception handling irregularity       & \\
Type casts           & Contract execution type-check error   & \\
Re-entrance flaw     & Function re-entered prior exit        & \\
Field disclosure     & Private value published by miner      & \\
\hline
Immutable bug        & Contract altering after deployment    & EVM \\
Ether lost           & Send ether to orphan address          & bytecode \\
\hline
Unpredicted state    & Contract state change prior call      & Blockchain \\
Randomness bug       & Seed biased by malicious miner        & mechanism \\
Timestamp failure    & Malicious miner alters timestamp      & \\
\hline
\end{tabular}
\end{table}

\section{Privacy and data protection aspects}
\label{sec:privacy}

Although blockchain is being considered as a somehow {\em anonymous data structure}, privacy properties in this context have never been formally stated in a provable way. Privacy can be considered as the right of an individual to control how personal information is obtained, processed, distributed, or used by others; i.e. it is related to personal data processing. The term {\em personal data} refers to information relating to an identified or identifiable natural person ---a person who can be identified, directly or indirectly. However, to determine whether a natural person is identifiable, account should be taken of all the means reasonably likely to be used for identifying the natural person directly or indirectly. Hence, personal data having undergone pseudonymisation that could be attributed to a natural person by using additional information, should be considered to be personal data.

In the case of incorporating blockchains in IoT technologies, the IoT devices will exchange data via the distributed ledger and smart contracts. In this scenario, each device can be singled out, roughly resulting in {\em device fingerprinting} as each device leaves a unique trace. Hence, if a device is associated with an individual, then personal data processing is in place. The above are in accordance with the European {\em general data protection regulation} (GDPR) EU 2016/679 stating that pseudonymisation should not be considered as anonymization, though it reduces the risks to the data subjects concerned. The GDPR is expected to apply to the majority of organisations, even if they lack establishments in the EU, if the subjects whose data are processed reside in the EU.

Thus, personal data protection issues in the blockchain are related to pseudonymisation and other privacy enhancing technologies that reduce privacy risks, like users' behaviour profiling without their consent. Clearly, the specific context of the blockchain under consideration is crucial for determining the associated  risks; these are higher in permissionless ledgers, where anyone can view the whole history of transactions. Several approaches for mitigating privacy issues have been proposed, where the majority concerns cryptocurrencies. However, these may also apply appropriately adjusted to blockchain for IoT security, as their goal is to avoid having user information revealed; monitoring users' activities for profiling purposes through automated decision-making tools is a typical example that poses significant risks for individuals' rights and freedoms.

The use of mixing schemes is a privacy enhancing approach where many users' transactions are mixed; As the need for a third party raises security issues, effort has been put to get mixing schemes operating in a transparent and verifiable way \cite{bonneau14}. In any case, the privacy obtained by such schemes needs to be evaluated since partial information leakage still occurs. Another approach rests with {\em zero knowledge} (ZK) proofs; the {\em ZK succinct non-interactive argument of knowledge} (zkSNARK) is a particular ZK proof that does not necessitate the interaction between provers and verifiers. This tool has been proposed for achieving anonymity in the blockchain ---with Zerocash system being an example. Its main idea is that a transaction's creator can prove that the transaction is true without revealing sender's or receiver's address and the transaction amount. A more recent approach is the design of a privacy preserving distributed~file storage system relying on the blockchain for handling funds with financial incentives given to storage providers \cite{kopp17}; a privacy preserving payment mechanism, based on ring signatures, and one-time addresses are at the core of system's design. Although the above approaches mainly target at the financial sector, the mathematical tools they are based on could be possibly applied, as stated above, to blockchain applications for the IoT industry; besides, such approaches have already been studied in many other frameworks, like e-voting and anonymous routing. In any case, it is clear that privacy issues are not fully resolved and further research is needed.

Another challenge that blockchain applications may need to address, to ensure compliance with the regulatory framework, is how to erase personal data from the ledger when a user revokes consent (if applicable) for such processing; this is referred to as the {\em right to be forgotten} in the GDPR. Towards this end, a number of solutions could be considered. For instance, the blockchain can contain the transactions' hash values and not the transactions themselves, which could be stored separately; deleting the separate transactions seems to address the right to be forgotten without affecting the overall solution.

\section{Current market situation}
\label{sec:market}

Several industry players have delivered blockchain solutions that aim at strengthening IoT. The partnership of IBM and Samsung  led to the {\em autonomous decentralised P2P telemetry} (ADEPT) platform. In particular, Ethereum was used for device coordination, delivering functions like registration, authentication and consensus-based blacklisting. Gladious recently proposed an approach for mitigating DDoS attacks using blockchain, where pools of nodes are dynamically formed (via Ethereum's smart contracts) to validate requested connections and block malicious activity. Other blockchain security tools for the IoT, like Factom, Filament, and Guardtime, have been developed focusing on safeguarding the integrity of system components.

The vast number of applications that can benefit from the blockchain leads to diverse requirements that cannot be met by a particular choice of DLT model or consensus protocol. The Hyperledger project, which is hosted by the Linux foundation, is a collaborative effort aiming at creating open-source DLT frameworks that will be the basis for building blockchain solutions. Amongst the developed frameworks, Hyperledger Fabric provides the components needed to address the heterogeneity of the IoT, meet user needs and easily deploy enterprise grade applications. Implementing a global ledger of public IoT devices requires first solving the scalability problem that is related to {\em the price of a public blockchain transaction and storage}. Private blockchains still have the advantage that they allow common management of a shared infrastructure with no need for a third-party.

\section{Conclusions}
\label{sec:conclude}

IoT devices have a reputation for being critically vulnerable with a collective power allowing them to impact targets beyond a typical attacks' scope. Blockchain seems to offer the tools needed for enhancing the security of IoT devices and address key challenges. The ability to define a framework for trusted transaction processing and coordination will allow IoT devices to communicate with the increased transparency and auditing. However, as blockchain products are being developed, compliance with the data privacy regulatory framework need also be taken into account, as it may affect important aspects of an envisaged solution.

\section{Acknowledgements}
\label{sec:thanks}

This work was supported by CYBER-TRUST project, which has received funding from the European Union's Horizon 2020 research and innovation programme under grant agreement no.\ 786698.

\section{About the authors}
\label{sec:authors}

Nicholas Kolokotronis (\href{mailto:nkolok@uop.gr}{nkolok@uop.gr}) is an Assistant Professor at the University of Peloponnese, Greece. His research interests include cryptography, security, and distributed ledger technologies.

Konstantinos Limniotis (\href{mailto:klimniotis@dpa.gr}{klimniotis@dpa.gr}) is an ICT auditor at the Hellenic Data Protection Authority, Greece. His research interests include cryptography and personal data protection.

Stavros Shiaeles (\href{mailto:stavros.shiaeles@plymouth.ac.uk}{stavros.shiaeles@plymouth.ac.uk}) is a Lecturer in cybersecurity at Plymouth University, UK. His interests include digital forensics, denial of service attacks, and intrusion detection.

Romain Griffiths (\href{mailto:romain.griffiths@neofacto.com}{romain.griffiths@neofacto.com}) is the CTO of Neofacto, France, and a technical advisor of Scorechain SA, Luxembourg, both companies specialising in Blockchain technologies.


\end{document}